\documentclass[preprint,review,12pt]{elsarticle}

\usepackage{amssymb}

\journal{Optics Communications}

\newcommand{\pra}{Phys. Rev. A\ }
\newcommand{\prd}{Phys. Rev. D\ }
\newcommand{\prl}{Phys. Rev. Lett.\ }
\newcommand{\pr}{Phys. Rev.\ }

\newcommand{\jpa}{J. Phys. A\ }

\newcommand{\etal}{{\em et al.}}

\newcommand{\oc}{Opt. Commun.\ }

\newcommand{\UQ}{School of Mathematics and Physics, University of Queensland, Brisbane, 
QLD 4072, Australia.}

\begin{document}

\begin{frontmatter}

\title{Improved quantum correlations in second harmonic generation with a squeezed pump}

\author[em]{E. Marcellina}
\author[jfc]{J.~F. Corney}
\author[mko]{M.~K. Olsen\corref{cor1}}
\ead{mko@physics.uq.edu.au}

\cortext[cor1]{Corresponding author}

\address{\UQ}

\begin{abstract}

We investigate the effects of a squeezed pump on the quantum properties and conversion efficiency of the light produced in single-pass second harmonic generation. Using stochastic integration of the two-mode equations of motion in the positive-P representation, we find that larger violations of continuous-variable harmonic entanglement criteria are available for lesser effective interaction strengths than with a coherent pump. This enhancement of the quantum properties also applies to violations of the Reid-Drummond inequalities used to demonstrate a harmonic version of the Einstein-Podolsky-Rosen paradox. We find that the conversion efficiency is largely unchanged except for very low pump intensities and high levels of squeezing.  

\end{abstract}


\end{frontmatter}

\section{Introduction}
\label{sec:intro}

One of the simplest non-linear optical processes is travelling wave second harmonic generation~\cite{DFW}, in which a nonlinear $\chi^{(2)}$ crystal is pumped with an electromagnetic field at one frequency and produces a second harmonic field at twice this frequency. A comprehensive classical treatment of this process was first given by Armstrong \etal~\cite{JAA}. The quantum properties of the output fields were first calculated using a method of linearisation about the classical solutions~\cite{linear}, even though it had long been known that these were not accurate for arbitrary interaction strengths~\cite{CTT}. The approach of Walls and Tindle, using matrix equations for the number state coefficients, was necessarily limited to small photon numbers. The later development of the positive-P representation~\cite{P+}, which maps the quantum evolution equations onto stochastic differential equations in a doubled phase space, allowed for the treatment of much larger photon numbers. For this system, this was first taken advantage of by Olsen \etal~\cite{revive} to treat the two-mode model, finding that full conversion to the second harmonic did not occur, but that the fundamental field experienced a revival inside the crystal. This approach enabled a calculation of the quantum properties, such as quadrature squeezing, of the output fields without relying on any assumptions about the mean-field solutions. The positive-P representation was also used to calculate the QND (Quantum Non-Demolition) properties of the system~\cite{QND1} and also compared with results found using the semi-classical method theory of stochastic electrodynamics~\cite{QND2}. The quantum correlations between the two fields were also calculated~\cite{sumdiff}, using correlations which later became famous as the Duan-Simon criteria for two-mode continuous-variable entanglement~\cite{Duan,Simon}. The entanglement between the fundamental and harmonic fields was later named harmonic entanglement by Grosse \etal~\cite{harmKoy}, who calculated it for a system which could operate in both the up and down-conversion regimes.
The Reid correlations~\cite{ReidEPR} for Einstein-Podolsky-Rosen entanglement~\cite{EPR} between the two modes have also been calculated previously~\cite{SHGEPR}. In the spirit of Grosse \etal, we shall name this harmonic steering.

The idea that the conversion efficiency in nonlinear optical processes could be a function of the quantum statistics of the inputs was raised by Shen~\cite{earlystat} in 1967. Shen showed that the conversion efficiency in the two-mode model of second harmonic generation would depend on the second-order correlation function, $g^{(2)}(0)$~\cite{Roy}, of the pump, predicting that light with chaotic statistics would initially convert twice as efficiently as a coherent pump.
This was later verified by stochastic integration, where the differences in efficiency between these pumps were explicitly calculated~\cite{intherm} without making the small interaction approximations required in Shen's approach.  The development of algorithms to model different quantum states in the positive-P and Wigner representations~\cite{Qstates} allowed for the investigation of the effects of these in the process of atom-molecule conversion in trapped Bose-EInstein condensates~\cite{BEC1,BEC2}, showing that the initial quantum statistics would also affect this process. Because squeezed states with low average photon number have $g^{(2)}(0)>1$, we decided to investigate the effects of input squeezing in this system.  In this work, we use the method given in Ref.~\cite{Qstates} to model the input to our crystal as squeezed states with different degrees of amplitude and phase quadrature squeezing, and investigate the outputs in terms of conversion efficiency, single-mode squeezing, two-mode entanglement, and the EPR paradox as expressed by the Reid criteria.

\section{Squeezed states}
\label{sec:squeezado}

Before we enter into the calculations for the problem, we will review the definition of squeezed states and the quadratures as we define them in this work. This is necessary as there are several different conventions found in the literature, which can lead to confusion if the quantities used are not explicitly defined. We begin with the bosonic annihilation operators, $\hat{a}$ for photons in the fundamental mode at frequency $\omega$, from which we define the quadrature operators
\begin{equation}
\hat{X}_{a}=\hat{a}+\hat{a}^{\dag},\hspace{1cm}\hat{Y}_{a}=-i\left(\hat{a}-\hat{a}^{\dag}\right),
\label{eq:quads}
\end{equation}
with similar definitions for the harmonic mode at $2\omega$, using the operator $\hat{b}$. With this definition of the quadratures, the vacuum or coherent state level of the variances is
\begin{equation}
V(\hat{X_{j}})=V(\hat{Y}_{j})=1,
\label{eq:vars}
\end{equation}
where $j=a,b$. 

A squeezed state is then a state of the electromagnetic field that has a variance of less than one in one of the quadratures, at the expense of increased fluctuations in the orthogonal quadrature. We note here that the quadratures can be defined at any angles, but that zero and $\pi/2$ are sufficient for our purposes here. A coherently displaced squeezed state is written as $|\alpha,r\mbox{e}^{i\phi}\rangle$, where the c-number $\alpha$ represents the coherent displacement and $r$ is known as the squeezing parameter, with $\phi$ the angle of the squeezed quadrature. This state has a mean intensity, $N_{a}=|\alpha|^{2}+\sinh^{2}r$ and quadrature variances for $\phi=0$, $V(\hat{X}_{a})=\mbox{e}^{-2r}$ and $V(\hat{Y}_{a})=\mbox{e}^{2r}$. For $\phi=\pi/2$, the $\hat{Y}$ quadrature is squeezed and $\hat{X}$ is anti-squeezed. The formal squeezing operator is $S(\epsilon)=\exp(1/2\epsilon^{\ast}\hat{a}^{2}-1/2\epsilon\hat{a}^{\dag\;2})$~\cite{squeezy}, where $\epsilon=r\mbox{e}^{2i\phi}$. As this operator creates and annihilates photons in pairs, we might expect a squeezed field to lead to increased second harmonic conversion efficiency, which we will investigate in what follows.

The second-order correlation function is
\begin{equation}
g^{(2)}(0)=\frac{\langle\hat{a}^{\dag}\hat{a}^{\dag}\hat{a}\hat{a}\rangle}{\langle\hat{a}^{\dag}\hat{a}\rangle^{2}}, 
\label{eq:gtwo}
\end{equation}
which can be expressed in terms of the number variance as
\begin{equation}
g^{(2)}(0) = 1+\frac{V(N)-\langle\hat{a}^{\dag}\hat{a}\rangle}{\langle\hat{a}^{\dag}\hat{a}\rangle^{2}}.
\label{eq:g2V}
\end{equation}
We can immediately see that if we have a super-Poissonian field with $V(N)>\langle\hat{a}^{\dag}\hat{a}\rangle$, $g^{(2)}(0)$ will be greater than unity and we will find photon bunching. On the other hand, $V(N)<\langle\hat{a}^{\dag}\hat{a}\rangle$ will give an anti-bunched field.

The explicit expressions for the number variances are found as~\cite{DFW}
\begin{eqnarray}
V(N) &=& |\alpha|^{2}\exp(-2r)+2\sinh^{2}r\cosh^{2}r, \hspace{1cm}\mbox{X squeezed}\nonumber\\
V(N) &=& |\alpha|^{2}\exp(2r)+2\sinh^{2}r\cosh^{2}r, \hspace{1cm}\mbox{Y squeezed},
\label{eq:VNXY}
\end{eqnarray}
from which we see that a field squeezed in the $\hat{X}$ quadrature will tend to exhibit anti-bunching while one squeezed in the $\hat{Y}$ quadrature will tend to exhibit bunching. However, calculations show that even for a coherent amplitude of $\alpha=10^{2}$, the values of $g^{(2)}(0)$ remain close to unity. As the coherent amplitude increases, they diverge even less from unity, so that we would not expect a squeezed pump to result in significantly enhanced conversion efficiency except for very weak pumps. We will investigate the effects on the quantum properties of the output fields below, using stochastic integration.

\section{Hamiltonian and Equations of Motion}
\label{sec:Ham}

In this work we will use a simplified description of travelling wave second harmonic generation which does not treat dispersion within the $\chi^{(2)}$ medium and treats the fields as plane waves at fixed frequencies, $\omega$ and $2\omega$. Whilst this approach would not be highly accurate if we sought to model an actual experiment, it is sufficient to show the effects we are looking for as functions of the degree of squeezing of the pump beam. The interaction Hamiltonian is written as
\begin{equation}
{\cal H} = i\hbar\frac{\kappa}{2}\left[\hat{a}^{\dag\;2}\hat{b}-\hat{a}^{2}\hat{b}^{\dag}\right],
\label{eq:Ham}
\end{equation}
where $\kappa$ is the effective nonlinearity. Following the standard procedures~\cite{QNGardiner}, we map this Hamiltonian onto a set of four stochastic differential equations for the evolution of the c-number variables of the positive-P phase-space representation as the fields traverse the nonlinear medium. Proceeding via the von Neumann and Fokker-Planck equations, we find
\begin{eqnarray}
\frac{d\alpha}{dz} &=& \kappa\alpha^{+}\beta+\sqrt{\kappa\beta}\eta_{1}(z),\nonumber\\
\frac{d\alpha^{+}}{dz} &=& \kappa\alpha\beta^{+}+\sqrt{\kappa\beta^{+}}\eta_{2}(z),\nonumber\\
\frac{d\beta}{dz} &=& -\frac{\kappa}{2}\alpha^{2},\nonumber\\
\frac{d\beta^{+}}{dz} &=& -\frac{\kappa}{2}\alpha^{+\;2},
\label{eq:PPeqns}
\end{eqnarray}
where the $\eta_{j}(z)$ are real Gaussian noise terms with the properties $\overline{\eta_{j}(z)}=0$ and $\overline{\eta_j (z)\eta_k (z')}=\delta_{jk}\delta (z-z')$. The independence of the two noise terms means that $\alpha$ and $\alpha^{+}$ are not complex conjugate except on average, which also holds for $\beta$ and $\beta^{+}$. It is this property which allows us to integrate what are equivalent on average to equations of motion for non-commuting operators. Averages over a large number of trajectories of the system of equations above allow us to find normally-ordered operator expectation values, with
\begin{equation}
\overline{\alpha^{m}\alpha^{+\;n} }\rightarrow \langle\hat{a}^{\dag\;n}\hat{a}^{m}\rangle,
\label{eq:correspondence}
\end{equation}
and similarly for $\hat{b},\:\hat{b}^{\dag}$ and $\beta,\:\beta^{+}$. In practice we integrate these equations in Matlab and average over at least $10^{7}$ trajectories. 

\section{Squeezing for Different Input States}
\label{sec:results}

We have investigated the outputs for pump fields which are squeezed in either the $\hat{X}$ (amplitude) or $\hat{Y}$ (phase) quadratures, for squeezing parameters, $r=0,.5,1$. The results are shown in the graphs as a function of the parametrised interaction length, $\zeta=\kappa z\sqrt{(|\alpha(0)|^{2}+\sinh^{2}r)/2}$. In all our calculations in this section, we use $\alpha(0)=10^{3}$ and $\kappa=10^{-2}$. For these input parameters, we find no significant difference in the conversion efficiencies, consistent with the fact that $g^{(2)}(0)$ stays very close to unity. We did find that for even smaller $\alpha(0)$, the efficiency did improve more with increasing $\hat{Y}$ squeezing, but as we are interested in the quantum properties of reasonably intense outputs, we have not shown these results here.

The first quantum correlations we investigate are the quadrature variances, since bright squeezing can have applications in quantum communication and and teleportation~\cite{communication,teleport}. Beginning with a coherent pump, the equations of motion derived from our Hamiltonian predict almost perfect squeezing in the fundamental after a certain interaction length, with a maximum of $50\%$ squeezing in the harmonic, with the $\hat{X}$ quadrature being squeezed in both cases~\cite{revive}. After obtaining a maximum of squeezing, both become antisqueezed as the fundamental mode revives inside the nonlinear medium.  

\begin{figure}
\begin{center}
\includegraphics[width=0.55\columnwidth]{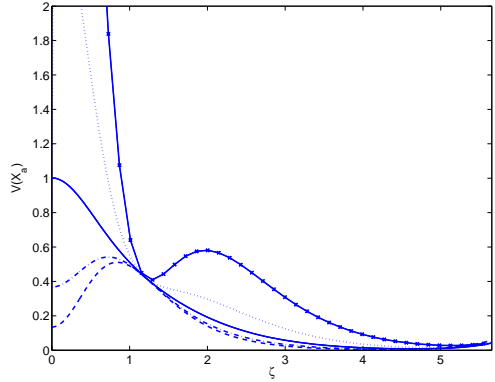}
\end{center}
\caption{$V(\hat{X}_{a})$ for different input states. On the left hand side of the plot, starting from the bottom, we show the values for $\hat{X}$-squeezed with $r=1$ (dashed line), $r=.5$ (dash-dotted line), $r=0$ (solid line), then $\hat{Y}$-squeezed with $r=.5$ (dotted line) and $r=1$ (dashed line with crosses). We note here that the method used to perform the integration does not allow for an easy determination of the sampling errors, but when we run the program several times and compare the results, we do not find significant differences. For the quantities shown here, the inferred errors would be of the size of the line widths. The quantities plotted in this and all subsequent plots are dimensionless.}
\label{fig:VXa}
\end{figure}

In Fig.~\ref{fig:VXa} we show the pump quadrature variances, $V(\hat{X}_{a})$, for different input statistics, with the solid line beginning at one being the prediction for a coherent pump.  We see that a pump squeezed in the $\hat{X}$ quadrature actually results in output fields which are squeezed to begin with, as expected, but that they lose some of this squeezing until they reach the same level as that for a coherent pump, which happens just beyond $\zeta=1$. After this they become increasingly squeezed, staying below the level for the coherent pump at least until the squeezing begins to diminish, at $\zeta \approx 4.5$. For a pump with $\hat{Y}$ squeezing, we see that the output fields begin as antisqueezed, with squeezing developing until they are equal to the other fields shown around $\zeta \approx 1.2$. The variance of the field resulting from the pump with $r=0.5$ continues to decrease, but stays above the level for the coherent input until $\zeta \approx 5$. Beyond the length where the variances are equal, it begins to increase for $r=1$, but then decreases again until the variances of the five different fields are indistinguishable at around $\zeta \approx 5$. This behaviour shows that the system has some memory of the input even when the fundamental field has reached its minimum and begun to revive, which happens in the region $4 \leq \zeta \leq 6$ for the parameters we use here~\cite{revive}. The $\hat{Y}_{a}$ quadrature always became antisqueezed, even when it began with $V(\hat{Y}_{a})<1$ as a result of a $\hat{Y}$ squeezed pump. The output quadrature was never more squeezed than the input quadrature, with the variances increasing monotonically over the range that we investigated.

\begin{figure}
\begin{center}
\includegraphics[width=0.55\columnwidth]{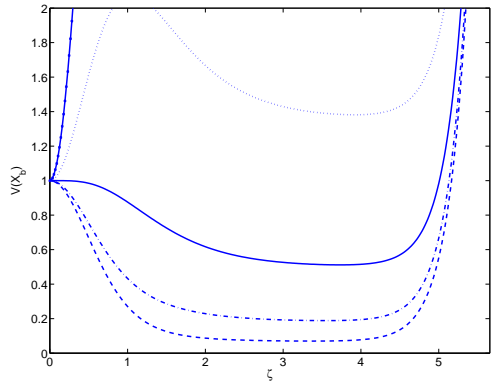}
\end{center}
\caption{$V(\hat{X}_{b})$ for different input states. On the left hand side of the plot, starting from the bottom, we show the values for $\hat{X}$-squeezed with $r=1$ (dashed line), $r=.5$ (dash-dotted line), $r=0$ (solid line), then $\hat{Y}$-squeezed with $r=.5$ (dotted line) and $r=1$ (uppermost dashed line with crosses).}
\label{fig:VXb}
\end{figure}

When we look at the variances in the harmonic, we find interesting behaviour, which is qualitatively different depending on which pump quadrature is squeezed. The effects on $V(\hat{X}_{b})$ of squeezed pumps are shown in Fig.~\ref{fig:VXb}. For a coherent pump we see that the squeezing in this output quadrature reaches a maximum at $V(\hat{X}_{b})\approx 0.5$, with $\hat{X}_{a}$-squeezed pumps resulting in a significantly greater degree of squeezing, while $\hat{Y}_{a}$-squeezing in the pump results in antisqueezing in the $\hat{X}_{b}$ output. However, when we look at the behaviour of the $\hat{Y}_{b}$ quadrature, we see that a $\hat{Y}_{a}$ (phase) squeezed pump leads to squeezing in this $\hat{Y}_{b}$ quadrature, as well as in $\hat{X}_{a}$. This is in stark contrast to an amplitude quadrature squeezed pump, which gives squeezing in both the output amplitude quadratures. This effect may be of interest experimentally for controlling the harmonic quadrature in which squeezing is available, as this changes from amplitude to phase quadrature as the squeezed pump quadrature changes, while leaving the fundamental squeezing in the $\hat{X}_{a}$ quadrature.

\begin{figure}
\begin{center}
\includegraphics[width=0.55\columnwidth]{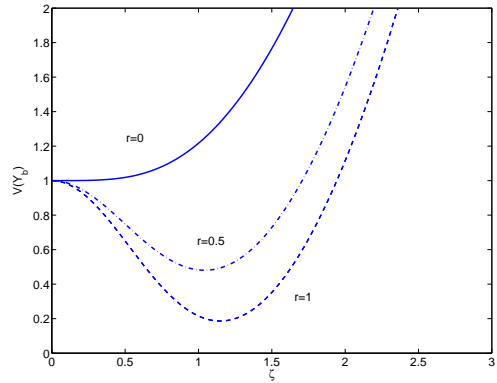}
\end{center}
\caption{$V(\hat{Y}_{b})$ for pump fields initially squeezed in the $\hat{Y}$ quadrature. $V(\hat{Y}_{a})$ begins at the initial pump values and rapidly becomes anti-squeezed. With an $\hat{X}$-squeezed pump, $V(\hat{Y}_{b})$ always shows noise above the coherent state level.}
\label{fig:VYb}
\end{figure}

\section{Entanglement and EPR for Different Input States}
\label{sec:DSEPR}

Squeezing in and of itself is a single-mode phenomenon, and even though it has several potential applications, the field becomes more interesting when we consider multi-mode correlations, defined by combining the natural modes of the problem in various manners. For the present system, this allows us to consider continuous-variable bipartite entanglement~\cite{Duan,Simon} and demonstrations of the Einstein-Podolsky-Rosen paradox~\cite{EPR}, more recently labelled steering~\cite{HMWvolante}. Asymmetric Gaussian steering has previously been theoretically analysed using an unbalanced nonlinear coupler~\cite{Sarahvolante} and experimentally produced using parametric downconversion~\cite{Natphotonics}. The phenomenon of entanglement between the fundamental and harmonic fields has been predicted~\cite{sumdiff} and subsequently labelled harmonic entanglement~\cite{harmKoy}. Following in this spirit, we will use the name harmonic EPR for the EPR correlations we calculate between the fundamental and harmonic fields.

We will use the Duan-Simon inequalities in their simplest forms, where
\begin{equation}
V(\hat{X}_{a}\pm\hat{X}_{b})+V(\hat{Y}_{a}\mp\hat{Y}_{b}) < 4
\label{eq:DSineq}
\end{equation}
guarantees that field $a$ and field $b$ are entangled. We note here that a value of this inequality greater than or equal to $4$ does not prove that entanglement is not present, as we have not taken advantage of the optimisation available in the original definition. This is not important for our purposes here, because we will later examine the EPR/steering correlations, which are of more practical significance for quantum information tasks. It is easily seen by inspection that two uncorrelated coherent states will give a value of $4$ for the expression in Eq.~\ref{eq:DSineq}.

\begin{figure}
\begin{center}
\includegraphics[width=0.55\columnwidth]{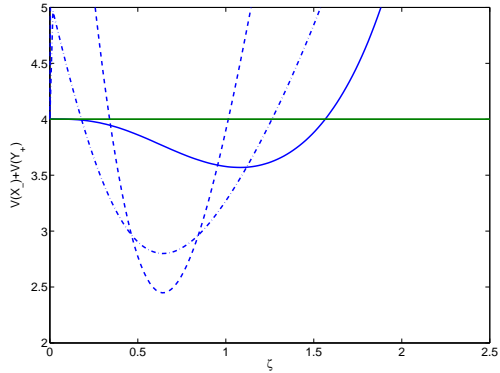}
\end{center}
\caption{The Duan-Simon correlations with the pump squeezed in the $\hat{X}$ quadrature. Solid line is $r=0$, dash-dotted line is $r=0.5$ and dashed line is $r=1$. The straight line gives the value expected for uncorrelated coherent states.}
\label{fig:XminusYplus}
\end{figure}

\begin{figure}
\begin{center}
\includegraphics[width=0.55\columnwidth]{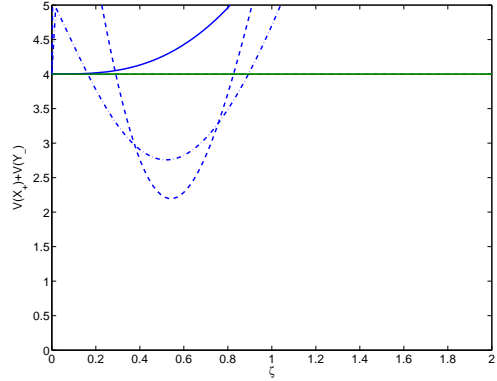}
\end{center}
\caption{The Duan-Simon correlations with the pump squeezed in the $\hat{Y}$-quadrature. Solid line is $r=0$, dash-dotted line is $r=0.5$ and dashed line is $r=1$. The straight line gives the value expected for uncorrelated coherent states.}
\label{fig:XplusYminus}
\end{figure}

In Fig.~\ref{fig:XminusYplus} we show the results obtained when the pump is $\hat{X}$-quadrature squeezed, with the solid line being for an initial coherent state with $r=0$, the dash-dotted line representing $r=0.5$, and the dashed line representing $r=1.0$. We readily see that increasing the input squeezing increases the degree of violation of the inequality, while also causing the violation to begin for smaller interactions. In this case, the violation is found by measuring the combined quadrature variance $V(\hat{X}_{a}-\hat{X}_{b})+V(\hat{Y}_{a}+\hat{Y}_{b})$.
In Fig.~\ref{fig:XplusYminus} we show the effects of input squeezing in the $\hat{Y}$-quadrature. We immediately see that in this case, the violation is found by measuring the combined quadrature variance $V(\hat{X}_{a}+\hat{X}_{b})+V(\hat{Y}_{a}-\hat{Y}_{b})$, which is not violated for a coherent pump. The lines for different squeezing values are as in Fig.~\ref{fig:XminusYplus}. Although not investigated here, it seems reasonable to infer that the violations will be found at different quadrature angles of the expression in Eq.~\ref{eq:DSineq} as a continuously changing and linear function of the angle of input squeezing. This tunability could possibly be of use in quantum information applications.

In order to investigate harmonic EPR/steering we have extended the approach of Reid~\cite{ReidEPR} to use two-frequency inferred inequalities as previously defined by Olsen~\cite{SHGEPR}. The basis of the EPR paradox is that, by inferring the values of one quadrature by using measurements on a conjugate quadrature, and then reversing the operation, we can obtain inferred measurements with a product of uncertainties which seemingly violates the Heisenberg Uncertainty Principle. Of course, as has been exhaustively discussed in the literature, there is no actual violation due to the fact that the quadrature values have no defined reality before the act of measurement. To see the paradox in action, we define inferred values of quadrature $\hat{X}_{a},\;\hat{Y}_{a}$ which may be obtained via measurements of quadrature $\hat{X}_{b},\;\hat{Y}_{b}$, and vice versa,
\begin{eqnarray}
V^{inf}(\hat{X}_{a}) &=& V(\hat{X}_{a})-\frac{\left[V(\hat{X}_{a},\hat{X}_{b})\right]^{2}}{V(\hat{X}_{b})},\nonumber\\
V^{inf}(\hat{Y}_{a}) &=& V(\hat{Y}_{a})-\frac{\left[V(\hat{Y}_{a},\hat{Y}_{b})\right]^{2}}{V(\hat{Y}_{b})},\nonumber\\
V^{inf}(\hat{X}_{b}) &=& V(\hat{X}_{b})-\frac{\left[V(\hat{X}_{a},\hat{X}_{b})\right]^{2}}{V(\hat{X}_{a})},\nonumber\\
V^{inf}(\hat{Y}_{b}) &=& V(\hat{Y}_{b})-\frac{\left[V(\hat{Y}_{a},\hat{Y}_{b})\right]^{2}}{V(\hat{Y}_{a})},
\label{eq:EPRdefs}
\end{eqnarray}
where $V(A,B)=\langle AB\rangle-\langle A\rangle\langle B\rangle$. We see a demonstration of the paradox whenever one of the following holds
\begin{eqnarray}
V^{inf}(\hat{X}_{a})V(\hat{Y}_{a}) &<& 1,\nonumber\\
V^{inf}(\hat{X}_{b})V^{inf}(\hat{Y}_{b}) &<& 1.
\label{eq:violate}
\end{eqnarray}
We note here that this is different to the standard definition, where quadrature measurements at the same frequencies are used. In our case, we are using a quadrature measurement at the fundamental (harmonic) frequency to infer values at the harmonic (fundamental) frequency. Since the Heisenberg Uncertainty Principle holds for quadrature variances at the same frequencies, satisfaction of the inequalities of Eq.~\ref{eq:violate} provides a demonstration of the paradox. A satisfaction of one of the inequalities but not the other would provide an example of asymmetric harmonic steering, but we did not find this here, with both the correlations of Eq.~\ref{eq:violate} returning identical values.

\begin{figure}
\begin{center}
\includegraphics[width=0.55\columnwidth]{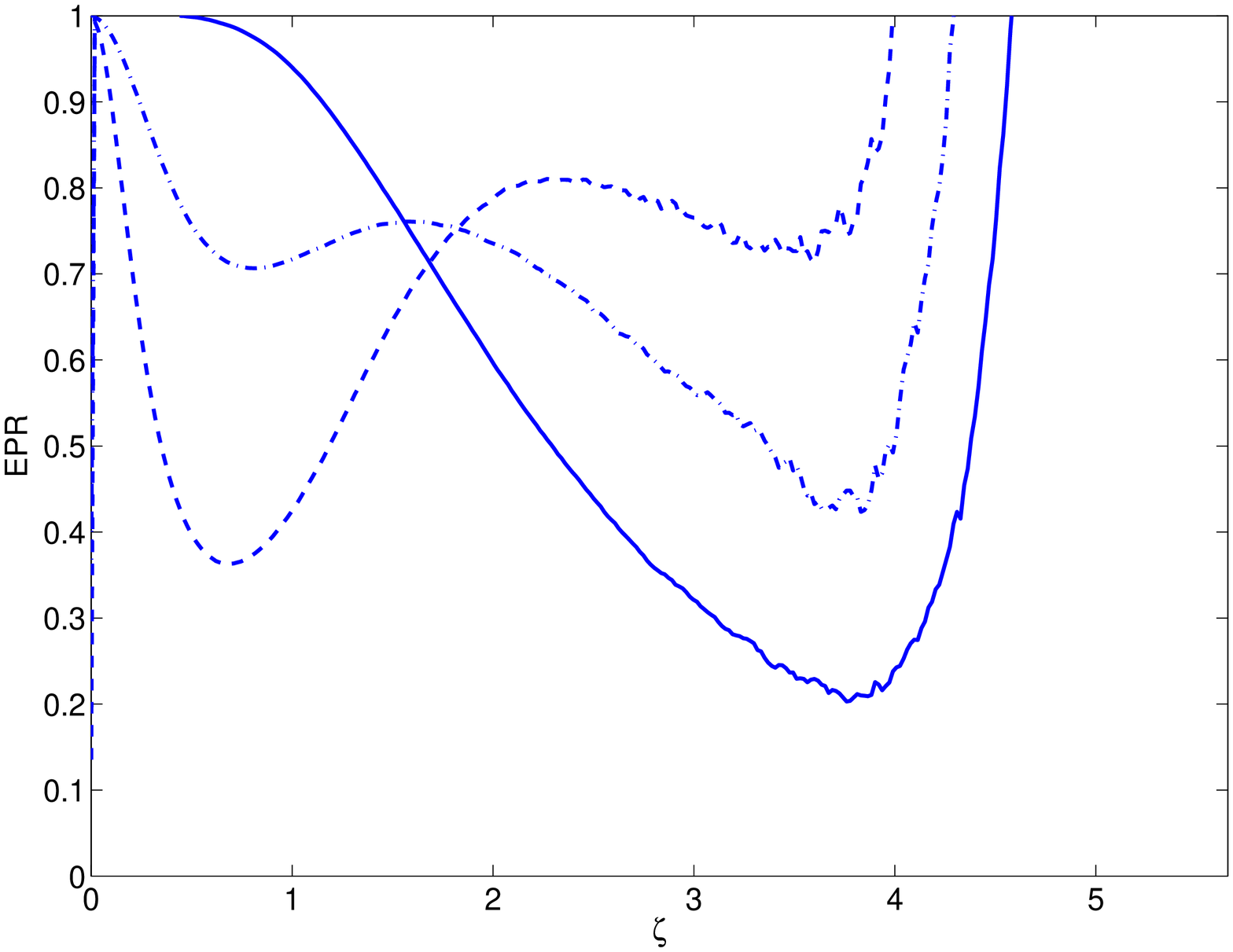}
\end{center}
\caption{The harmonic EPR correlation $V^{inf}(\hat{X}_{a})V^{inf}(\hat{Y}_{a})$ for an input pump squeezed in the $\hat{X}$-quadrature. We note that the sampling error has increased in the region where the denominators obtain their minima, for the same numbers of trajectories as in previous figures.}
\label{fig:EPRX}
\end{figure}

\begin{figure}
\begin{center}
\includegraphics[width=0.55\columnwidth]{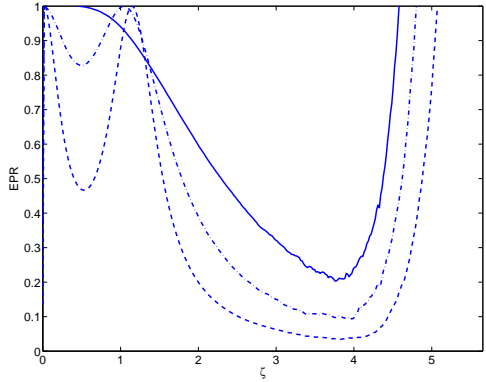}
\end{center}
\caption{The harmonic EPR correlation $V^{inf}(\hat{X}_{b})V^{inf}(\hat{Y}_{b})$ for an input pump squeezed in the $\hat{Y}$-quadrature.}
\label{fig:EPRY}
\end{figure}

In Fig.~\ref{fig:EPRX} we show the results for $V^{inf}(\hat{X}_{a})V^{inf}(\hat{Y}_{a})$ when we squeeze the input field in the $\hat{X}$-quadrature. The continuous line is for a coherent pump, while the dash-dotted line is for a pump with $r=0.5$ and the dashed line is for $r=1$. Fig.~\ref{fig:EPRY} is the same except for the fact that the pump is now squeezed in the $\hat{Y}$-quadrature. We see that squeezing the pump in the $\hat{X}$-quadrature gives a smaller maximum degree of the EPR correlation,  although the values for smaller interaction strengths are increased. This may be of some benefit experimentally both because this is the region where we would expect our simulations to be more realistic and because they may be obtained with a shorter nonlinear crystal or less intense pump. In Fig.~\ref{fig:EPRY} we see that $\hat{Y}$ squeezing gives stronger steering at both shorter and longer interaction strengths, with only a small region where a coherent pump is superior. The maximum of steering also increases as the pump is squeezed more. This suggests that, if we want increased harmonic steering from this system, a $\hat{Y}$-squeezed pump is superior apart from a small region. Since for a given crystal, the pump intensity can be tuned to stay away from this region, our results show clearly that pumping with a field squeezed in the $\hat{Y}$-quadrature is the best option. We note that the sampling errors increase in the regions where the denominators of the expressions are around their respective minima, but we have kept the same number of trajectories because the scaling of the errors changings with the square root of this number, meaning that we would need $10^{9}$ to see one tenth of the error. The time required would be prohibitive.

\section{Conclusions and Discussion}
\label{sec:conclude}

We have used a simple model of single-pass travelling wave second harmonic generation with a squeezed pump to invetsigate the effects of the squeezing on the quantum correlations of the outputs. We found that for the reasonably intense pumping that we have used in our model, the conversion efficiency is basically unchanged by pump squeezing, with the second order correlation function of the input only rising marginally above that of a coherent state. When we investigate the quadrature variances of the output fields, we find that a squeezed pump can increase the squeezing available in the harmonic field drastically, while having less of an effect on the fundamental.

The results are interesting when we look at the correlations between the fields at the two frequencies, in terms of both two-mode entanglement and EPR/steering. We find that use of a squeezed pump can increase the quantum correlations in the outputs, as well as making them available at lower interaction strengths. This could lead to greater flexibility in the use of second harmonic generation, possibly the simplest nonlinear optical process, as a valuable tool in quantum information and communication science and technology.

\section*{Acknowledgments}

This research was supported by the Australian Research Council under the Future Fellowships scheme. Elizabeth Marcellina was a visiting summer student with a scholarship from the School of Mathematics and Physics.

\end{document}